\documentclass[natbib,sn-basic]{sn-jnl}

\usepackage{graphicx} 

\jyear{2021}%

\theoremstyle{thmstyleone}%
\theoremstyle{thmstyletwo}%

\theoremstyle{thmstylethree}%

\usepackage{xcolor}

\begin{document}

\title{Understanding Active Region Emergence and Origins on the Sun and Other Cool Stars}


\author*[1]{\fnm{Maria A.} \sur{Weber}}\email{mweber@deltastate.edu}
\equalcont{These authors contributed equally to this work.}

\author[2]{\fnm{Hannah} \sur{Schunker}}\email{hannah.schunker@newcastle.edu.au}
\equalcont{These authors contributed equally to this work.}

\author[3]{\fnm{Laur\`ene} \sur{Jouve}}\email{ljouve@irap.omp.eu}
\equalcont{These authors contributed equally to this work.}

\author[4]{\fnm{Emre} \sur{I\c{s}{\i}k}}\email{emre.isik@tau.edu.tr}
\equalcont{These authors contributed equally to this work.}

\affil[1]{\orgdiv{Division of Mathematics and Sciences}, \orgname{Delta State University}, \orgaddress{\street{1003 W Sunflower Rd}, \city{Cleveland}, \postcode{38733}, \state{Mississippi}, \country{United States}}}

\affil[2]{\orgdiv{School of Information and Physical Sciences}, \orgname{University of Newcastle}, \orgaddress{\street{University Drive}, \city{Callaghan}, \postcode{2308}, \state{NSW}, \country{Australia}}}

\affil[3]{\orgdiv{IRAP/OMP/CNRS}, \orgname{Universit\'e Toulouse 3}, \orgaddress{\street{14 Avenue Edouard Belin}, \city{Toulouse}, \postcode{31400}, \country{France}}}

\affil[4]{\orgdiv{Department of Computer Science}, \orgname{Turkish-German University}, \orgaddress{\street{\c{S}ahinkaya Cd. 94}, \city{Beykoz}, \postcode{34800}, \state{Istanbul}, \country{Turkey}}}

\abstract{The emergence of active regions on the Sun is an integral feature of the solar dynamo mechanism. However, details about the generation of active-region-scale magnetism and the journey of this magnetic flux to the photosphere are still in question. Shifting paradigms are now developing for the source depth of the Sun's large-scale magnetism, the organization of this magnetism into fibril flux tubes, and the role of convection in shaping active-region observables. Here we review the landscape of flux emergence theories and simulations, highlight the role flux emergence plays in the global dynamo process, and make connections between flux emergence on the Sun and other cool stars. As longer-term and higher fidelity observations of both solar active regions and their associated flows are amassed, it is now possible to place new constraints on models of emerging flux. We discuss the outcomes of statistical studies which provide observational evidence that flux emergence may be a more passive process (at least in the upper convection zone); dominated to a greater extent by the influence of convection and to a lesser extent by buoyancy and the Coriolis force acting on rising magnetic flux tubes than previously thought. We also discuss how the relationship between stellar rotation, fractional convection zone depth, and magnetic activity on other stars can help us better understand flux emergence processes. 
Looking forward, we identify open questions regarding magnetic flux emergence that we anticipate can be addressed in the next decade with further observations and simulations.}

\keywords{Sun, solar, sunspot, magnetic field, flux emergence}



\maketitle


\section{Introduction}\label{sec1}

The Sun is a magnetically active star showing activity on a wide range of spatial scales and field strengths. An active region is defined by the appearance of a dark feature at the surface of the Sun in continuum white light observations. These features are associated with concentrations of strong magnetic fields, and often develop into fully formed, stable sunspots.  Typical active regions consist of opposite polarity pairs that are predominantly east-west aligned and have sizes ranging on the order of 10s to 100s of microhemispheres with lifetimes ranging from two days up to many weeks.

The Sun's coherent surface flux elements such as sunspots and active regions emerge from the solar interior. However, how they arrive at the surface and their specific depth of origin is not clear. Helioseismology has placed upper bounds on the amplitude and speed of the flows at and below the surface prior to emergence \citep[e.g.][]{Birchetal2013}, however any unambiguous detection of flows above the background noise remains a challenge. Therefore, numerical simulations of flux emergence through the surface of the Sun are critical to reconciling the observations with the physics of the formation of active regions.

Originally, the paradigm of an idealized, magnetically isolated flux tube was invoked to model magnetism giving rise to active regions. Here, it is assumed that the dynamo has already managed to create magnetism at the base of the convection zone a priori. Studies employing this paradigm were conducted first in a 1D Lagrangian frame using the thin flux tube approximation \citep[e.g.][]{Spruit1981,Caligarietal1995,Fanetal1993}, followed by 2D \citep[e.g.][]{MorenoInsertis1996,Fanetal1998} and 3D magnetohydrodynamic (MHD) \citep[e.g.][]{Abbettetal2001,Fan2008} approaches to resolve the flux tube cross-section and twist of magnetic field lines. As a body of work, these simulations suggest that magnetic buoyancy, the Coriolis force, and the twist of magnetic field lines in a tube play roles in the flux emergence process and are responsible for many active region observables. Addition of a convection velocity field further demonstrated that turbulent interior flows modulate flux emergence, provided the magnetic field strength of the flux tube is not substantially super-equipartition \citep[e.g.][]{Fanetal2003,JouveBrun2009,Weber2011}. However, this paradigm of idealized flux tubes built by a deep-seated dynamo mechanism has been challenged by results from 3D global convective dynamo simulations. Some demonstrate that toroidal wreaths of magnetism can be formed within the bulk of a stellar convection zone \citep[e.g.][]{Brownetal2011,Nelsonetal2011,Augustsonetal2015,MatilskyToomre2020}. Either from these wreaths \citep{Nelsonetal2011,Nelsonetal2013} or within the magneto-convection itself \citep{FanFang2014}, buoyantly rising magnetic structures - possibly starspot progenitors - are spawned. Results from both idealized flux tube simulations and the buoyant magnetic structures built self-consistently by dynamo action show similarities to active region observables, but there are also many discrepancies. Further modeling work with direct comparison to active region observations is critical to elucidate the true origin of active-region-scale magnetism.

The paradigm of the idealized, isolated flux tube mechanism for producing active regions is thus now changing towards a more complete picture. A large part of the recent paradigm shift was brought about from a statistical analysis of the flows associated with emerging active regions \citep[e.g.][]{Schunkeretal2016,Birchetal2019}, emphasising the importance of solar monitoring missions. Prior to instruments such as Helioseismic and Magnetic Imager (HMI) onboard NASA's Solar Dynamics Observatory (SDO) \citep{Scherreretal2012}, with high duty-cycle observations of the magnetic field and Doppler velocity at a cadence sufficiently shorter than the time it takes an active region to emerge, it was not possible to gain any statistical understanding of the emergence process to such detail. Similarly, monitoring campaigns for stars e.g., Mt Wilson, \emph{Kepler}, BCool, LAMOST, and TESS have increased both the sample size and the time range of data, such that magnetic variability has been measured over multiple cycle periods on other stars. Although the level of precision and sampling rate in such measurements are insufficient to amass emergence statistics for other stars like we have for the Sun, they help to shape our view over general trends of active-region formation in longitude and latitude, as well as the lifetimes of surface magnetic structures. 

The longest record of the eleven-year activity cycle of the Sun is defined by the number of sunspots, or cool, dark regions, visible on the Sun. Active regions are defined from this visible darkening when they are assigned an active region number by the National Oceanic and Atmospheric Administration. At the beginning of the solar cycle, sunspots appear at latitudes around $30^\circ$, and closer to the equator towards the end of the cycle, creating the observed butterfly diagram. Besides simply defining the solar cycle, active regions are found to have characteristics which correlate with the next solar cycle, suggesting that they are an integral part of the dynamo process. For example, the average tilt angle of sunspot groups over a solar cycle is anti-correlated with the amplitude of the next solar cycle \citep{Dasietal2010,Jiao+21}, and large active regions that emerge across the equator \citep[e.g.][]{Nagyetal2017} have a significant effect on the amplitude and duration of the subsequent solar cycle. Thus, to fully understand the dynamo process, it is critical to understand how active regions form.

Presumably the distribution of active regions at the surface of the Sun reflects the distribution of the global toroidal field in the interior \citep{Isik+11}, and can provide a strong constraint for their origin and the solar dynamo \citep[e.g.][]{Cameronetal2018}.
However, it cannot be excluded that the dynamo also produces strong field at latitudes which do not become unstable and rise to the surface. 
For other cool stars, the combined effects of rotation rate and fractional depth of the convection zone can lead to a possible mismatch between active regions on the surface and distributions of magnetic flux in the deeper interior due to latitudinal deflection as bundles of magnetism rise. As a result, any one-to-one association of observed surface field and the underlying dynamo in active cool stars is not necessarily straightforward \citep{Isik+11}.
While it is not currently possible to directly observe the emergence of a starspot, it is possible to make proxy observations \citep[e.g. from chromospheric indices, spectropolarimetry, Zeeman-Doppler imaging and asteroseismology;][]{Berdyugina2005,Garciaetal2010,Seeetal2016} to infer the  distribution, size, lifetime and magnetic field strength.

In this paper, we attempt to paint a comprehensive picture of the flux emergence process from generation of the active-region-scale magnetism in the deep interior to its appearance on the photosphere. We begin in Section \ref{sec:obs}, where we describe observations of the formation of active regions on the solar surface. These observations serve as inspiration and constraints for models of the generation and rise of emerging flux, which we review in Section \ref{sec:models}. New observations are highlighted in Section \ref{sec:nearsurface}, which support a more passive process for active region emergence than was previously understood based on flux emergence models. We then briefly review the role flux emergence plays in the solar dynamo process in Section \ref{sec:dynamo}, and discuss flux emergence leading to starspots on other cool stars in Section \ref{sec:stars}. In Section \ref{sec:forward}, we conclude with some recommendations as we move toward solving the active-region-scale flux emergence puzzle.

\section{Formation of active regions at the surface of the Sun}\label{sec:obs}
 
Active regions are defined by the appearance of dark spots on the visible disk of the Sun in white light, caused by strong, concentrated magnetic fields. The presence of this magnetism renders the spots cooler, and therefore darker, than the surrounding photosphere. Active region magnetic fields consist of roughly east-west aligned opposite polarity pairs, ranging from 10 up to 3000 micro-hemispheres in size, and $10^{20}$ to $10^{22}$~Mx of magnetic flux. Known as Hale's Law, bipolar active regions typically emerge with the same leading polarity in the same hemisphere, with the polarity orientation flipped for the opposite hemisphere \citep{Haleetal1919}. At the end of each 11-year sunspot cycle, the polarity orientation reverses for each hemisphere. In either hemisphere, active regions are roughly confined in toroidal bands which appear at higher latitudes of $\sim35^{\circ}$ at the beginning of each cycle and progressively move toward the equator over the roughly eleven year cycle.

\begin{figure}[ht]%
\centering
\includegraphics[width=0.99\columnwidth]{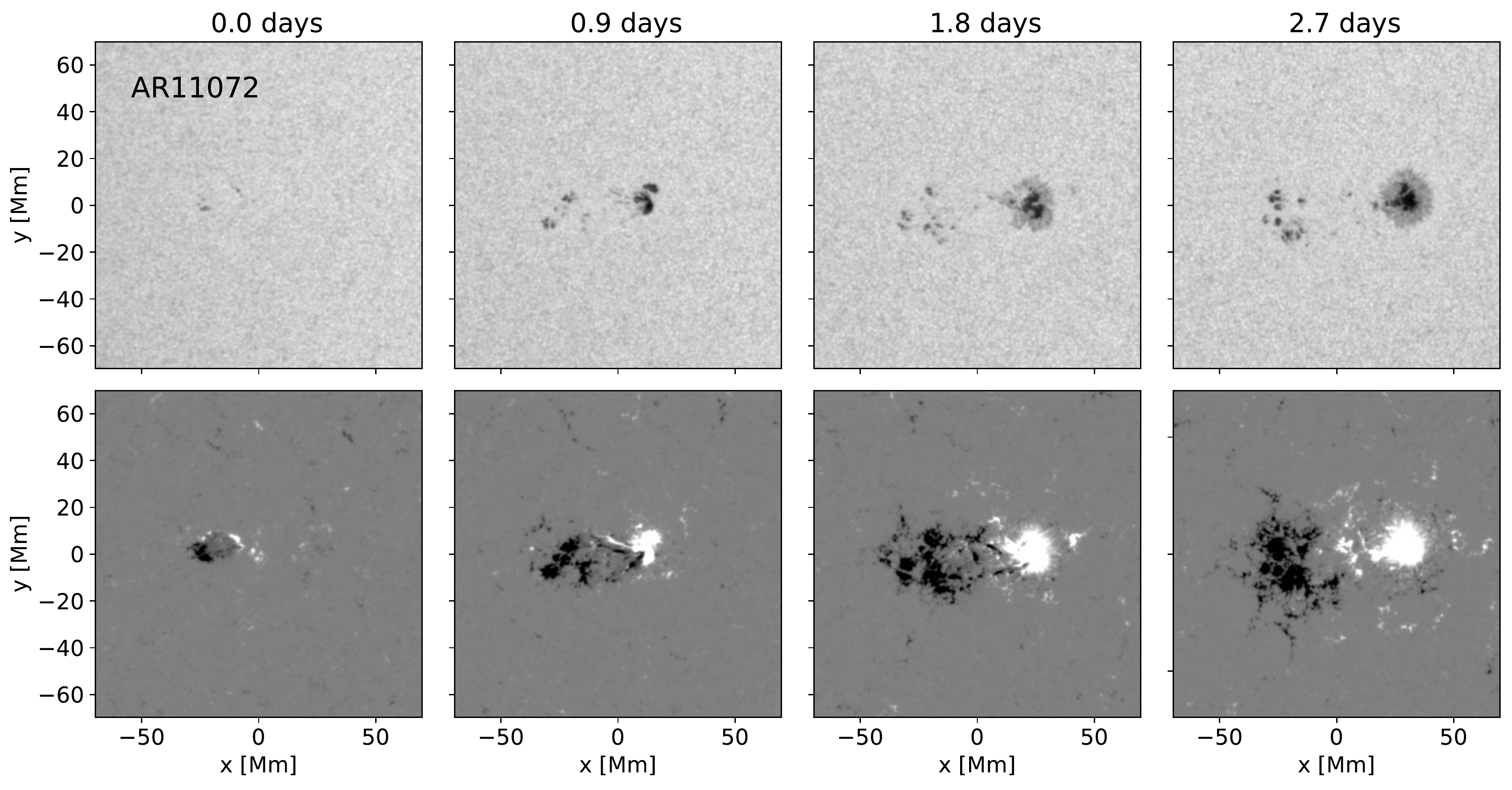}
\caption{Example of a typical active region, NOAA~AR11072, emerging onto the surface of the Sun as observed by SDO/HMI. The top row shows Postel projected maps of the continuum intensity, and the bottom row shows maps of the line-of-sight magnetic field $\pm 500$~G. In this instance, 0~days corresponds to the emergence time 2010.05.20\_17:12:00\_TAI, and the maps are centred at Carrington longitude $316.43^\circ$ and latitude $-15.13^\circ$. The east-west direction is $x$ and the north-south is $y$.
Hale's Law, the formation of a sunspot in the leading polarity, and Joy's Law are evident.}
\label{fig:ar11072}
\end{figure}

The leading polarity of an active region (in the prograde direction) also tends to be closer to the equator than the following polarity (in the retrograde direction). This statistical feature is known as Joy's Law \citep{Haleetal1919}. Joy's Law is often quantified by the `tilt angle' of the line drawn between the centers of leading and following polarity regions with respect to the east-west direction. Figure~\ref{fig:ar11072} shows the bipolar nature of a typical active region and illustrates Joy's Law, as the leading polarity of this southern-hemisphere active region is tilted closer to the equator.

An active region develops from a small magnetic bipole and grows in  size as more and more magnetic flux emerges (e.g. Fig.~\ref{fig:ar11072}). The flux-weighted centres of the polarities move further apart, predominantly in the east-west direction during the emergence process, as more flux emerges. The line-of-sight magnetic field observations show that magnetic field typically emerges as small scale features near the flux-weighted centre of the active region, which then stream towards the main polarities.
Active regions have lifetimes on the order of days to weeks, where large, high-flux active regions live longer than small, low-flux regions \citep[e.g.][]{SchrijverZwaan2000}.
Within the active regions, sunspots can form with peak magnetic field strengths from 2000 to 4000~G. Generally, the leading-polarity spot of the bipolar pair is larger and more coherent than the trailing-polarity region (see also Fig.~\ref{fig:ar11072}). Active regions also have a preferred hemispheric sense of magnetic helicity, as obtained from vector magnetograms. The observations favor a left-handed (negative) twist of the field lines in the northern hemisphere, and a right-handed (positive) twist in the southern hemisphere \citep[e.g.][]{Pevtsovetal2001,Pevtsovetal2003,Pevtsovetal2014,Prabhuetal2020}.

Having said how active regions \emph{typically} form, there is a wide variation in characteristics. When two or more polarity pairs emerge in the vicinity of one another, the polarities can morph into the traditional bipolar structure during the emergence process, usually leading to a more complex, multi-spot active region \citep[e.g. AR~11158 in][Fig.~1]{Schunkeretal2019}. It is also common to find active regions emerging into sites of existing magnetic field from previous active regions, so-called `nests' of activity \citep{Castenmilleretal1986}, where the emerging magnetic field interacts via cancellation and superposition with the existing magnetic field. 

\begin{figure}[ht]%
\centering
\includegraphics[width=0.9\columnwidth]{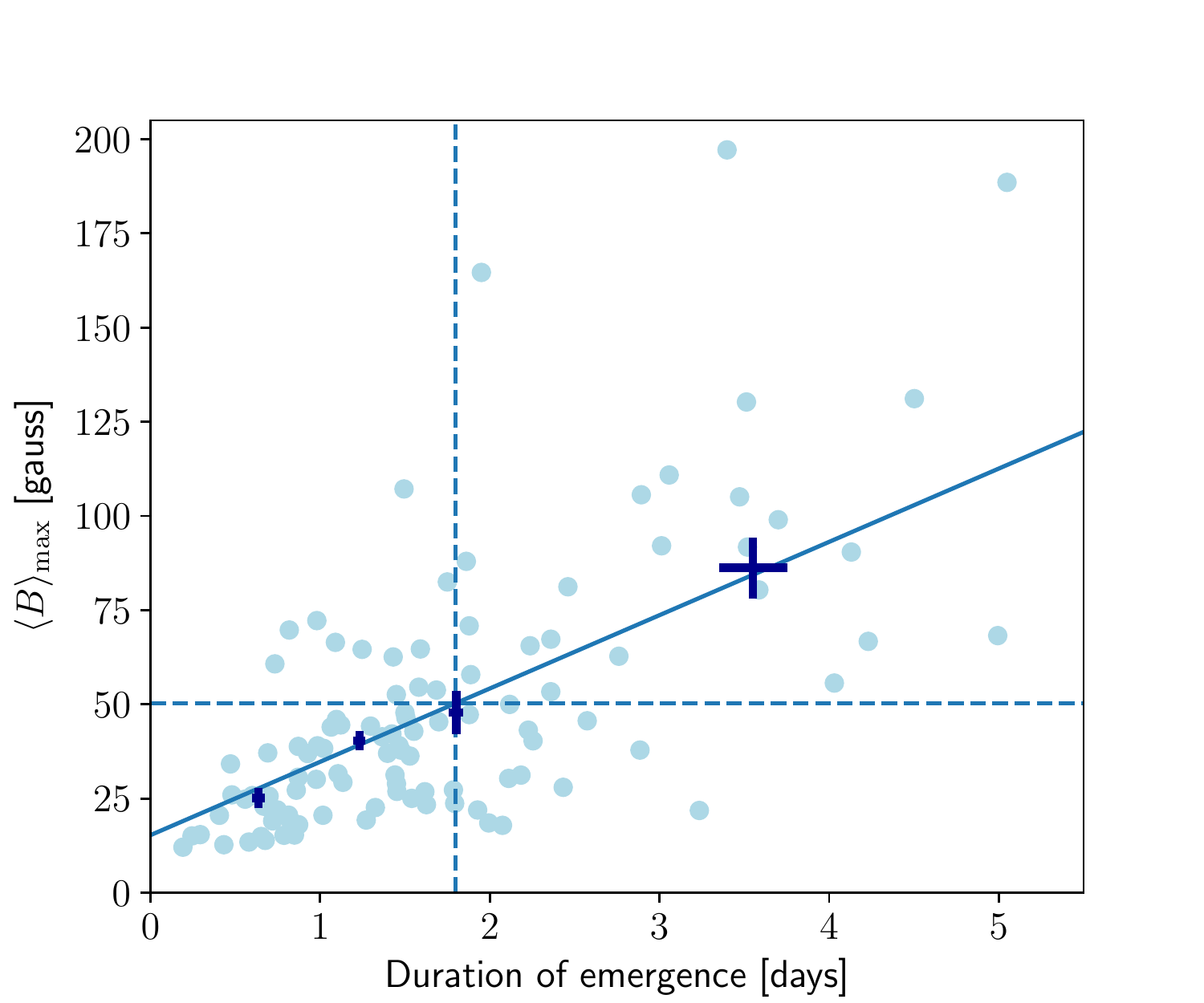}
\caption{Duration of the emergence process as a function of $\langle |B| \rangle_\mathrm{max}$ for each emerging active region (light blue points) and the blue diagonal line is a linear fit with slope $19.5 \pm 2.2$~G/day.  The blue horizontal dashed line is the mean of $\langle |B| \rangle_\mathrm{max}$,  $48.6 \pm  2.9$~G, and the vertical dashed line is the mean duration of emergence time $1.7 \pm 0.1$~days. The mean and uncertainty values of $\langle |B| \rangle_\mathrm{max}$ bins with equal number of points are shown in dark blue. See Appendix~\ref{app:earssepvel} for details on how the emergence time was calculated. }
\label{fig:emergtime}
\end{figure}

Figure~\ref{fig:emergtime} shows that the duration of the emergence process until magnetic flux has stopped increasing is, on average, linearly proportional to the maximum mean magnetic field  $\langle B \rangle_\mathrm{max}$ (see Appendix~\ref{app:earssepvel} for details on how the emergence time was calculated).
Given that the maximum flux of an active region is known to directly correlate with the lifetime \citep[e.g.][]{SchrijverZwaan2000}, our results are consistent with \cite{Harvey1993PhDT} (Chapter 3, Table III). Those results show that the ``rise time" of active regions with a smaller maximum area is 1-2 days and increases to 3-4 days for active regions with larger maximum area \citep{Harvey1993PhDT}. Here, we specifically avoid the term ``rise time" since it implies a physical rising. What we are actually measuring is the time it takes magnetic flux to stop increasing in an active region at the surface. Figure~\ref{fig:emergtime} shows that the relationship is, though with considerable scatter, remarkably linear, with a slope of $19.5 \pm 2.2$~G per day.

\section{Models of emerging flux}
\label{sec:models}

If the active-region-scale magnetism described in Section \ref{sec:obs} is generated by the underlying dynamo, then it must somehow make its way from the subsurface large-scale magnetic field to the surface. The appearance of active regions evokes the idea of rising ropes of magnetism. We see arches of magnetic bundles extended above the Sun’s surface. At the footpoint of these are sunspots. Within the Sun’s interior is where we think these bundles of magnetic flux are born, which then rise and intersect with the photosphere to form sunspots. In this section, we briefly review models and their outcomes which describe the formation of active-region-scale magnetic structures and their rise to the photosphere (also see the reviews by \cite{FanLRSP2021} and by \cite{CheungIsobe2014}).

\subsection{Formation and Destabilization of Active-Region-Scale Magnetic Structures}\label{sec:formB}
The magnetism responsible for active regions is formed in the solar interior, however, the exact physical location of magnetic field generation is not known with certainty. The paradigm that solar physics has clung on to is that the magnetism giving rise to active regions is generated and stored at the base of the convection zone in the weakly subadiabatic overshoot region \citep[e.g.][]{Parker1975,vanBallegooijen1982,MorenoInsertisetal1992,Rempel2003}. Here it is thought that shear from differential rotation at the tachocline transforms poloidal field into toroidal field, which is amplified until it is strong enough to become buoyantly unstable. Then the magnetism subsequently rises through the convection zone to the photosphere. Beyond this shearing and storage mechanism, many studies of flux emergence, assuming the magnetism is formed as `flux tubes' in the overshoot layer or at the very bottom of the convection zone, reproduce many properties of solar active regions (see \ref{sec:FluxTube}).

Studies have been carried out which consider magnetic buoyancy instabilities as a means to initiate the rise of magnetic flux bundles from the overshoot region \citep[e.g.][]{Spruit+1982,Spruit+1982erratum,FerrizMas+1995,Caligarietal1995}. Magnetic buoyancy is the result of a buoyant force due to the presence of a concentration of magnetism. Imagining this magnetism as a bundle or `tube' of magnetic flux, there is a pressure balance between the gas pressure outside the tube ($P_{e}$) and the sum of the gas pressure ($P_{i}$) and magnetic pressure ($P_{b}$) inside. The gas density of the tube can be reduced if there is a condition of temperature equilibrium, allowing the tube to buoyantly rise. Even if the tube is in neutral density with its surroundings, a perturbation could result in an undular instability that lifts part of the tube upward creating an ${\Omega}$-shaped loop, allowing mass to locally drain down the legs of the rising loop apex and initiating a buoyant rise. When considering thin flux tubes in mechanical equilibrium, their stability is primarily determined by their magnetic field strength and the subadiabaticity of the overshoot region \citep[e.g.][]{Caligarietal1995}. It is found that the field strength of the flux tube must exceed the equipartition value of $\sim10^4$ G by about an order of magnitude in order to develop unstable modes at sunspot latitudes in less than $\sim$1 year. 

Instead of considering isolated magnetic flux tubes built in the tachocline region, many studies using multi-dimensional MHD simulations have focused on the formation of buoyant instabilities within layers of uniform, horizontal magnetic field. \citep[e.g.][]{CattaneoHughes1988,Matthews+1995,Fan2001,VasilBrummell2008,VasilBrummell2009}. Indeed, it has been shown that regions of velocity shear can generate tube-like magnetic structures or magnetic layers \citep[e.g.][]{Clineetal2003,VasilBrummell2008}). \cite{VasilBrummell2008} find that a velocity shear representing a tachocline-like shear can generate a strong layer of horizontal magnetic field. From this self-consistently generated magnetic layer, buoyant structures resembling undulating `tubes' arise due to magnetic buoyancy instabilities within the magnetic layer. However, the shear required to develop the magnetic buoyancy instabilities of the magnetic layer is much stronger and the magnetic Prandtl number is much larger than what is expected in the solar tachocline \citep{VasilBrummell2009}. In order to generate a twist of the magnetic field within such rising magnetic `flux tubes', as is found in active region observations, \cite{Favier2012} showed that it was sufficient to add an inclined uniform weak field  on top of the unstable horizontal magnetic layer. Indeed, in this case, the unstable undulating tubes interact with the overarching inclined field as they buoyantly rise and the field lines start to wind around the tube axis, creating an effective twist in the magnetic structure.

There is now a shifting paradigm regarding the location of active-region-scale magnetic field generation. Recent global 3D magnetohydrodynamic (MHD) dynamo simulations have compelling outcomes which suggest that active-region-scale magnetism need not be formed at the base of the convection zone. In some, cyclic wreaths of magnetism are built amid the magneto-convection without the need for a tachocline \citep[e.g.][]{Brownetal2011,Augustsonetal2015}. Taking similar simulations but reducing sub-grid-scale turbulent diffusion \cite{Nelsonetal2011,Nelsonetal2013,Nelsonetal2014} capture the generation of buoyant magnetic structures arising from magnetic wreaths – possible starspot progenitors. While typical azimuthal field strengths are a few kilogauss, buoyant loops only spawned in regions with super-equipartition localized fields. The global dynamo simulations of \cite{FanFang2014} also exhibit super-equipartition flux bundles that rise toward the simulation upper domain. A common trait of these dynamo-generated buoyant loops is that they are continually amplified by shear and differential rotation as they rise. Unlike flux tube simulations (see \ref{sec:FluxTube}), these are not isolated  magnetic structures.  Yet recently, \cite{BiceToomre2022} found self-consistently generated flux ropes in a global 3D-MHD dynamo simulation representative of an early M-dwarf with a tachocline. The majority of the ropes remain embedded in the tachocline, while buoyant portions are lifted upward by nests of convection.

Taken together, these models and simulations of the formation and instability of buoyant magnetic structures, possible starspot progenitors, ask us to consider/reconsider the paradigm of isolated magnetic flux tubes arising from the deep convection zone. However, as is the case with all simulations, it is important to note that all the simulations discussed here are far removed from the regime of real stars. Yet, they reproduce the observed properties of active regions remarkably well and give us a glimpse into the complex interplay of forces and mechanisms at work in stellar interiors that conspire to generate magnetic structures and facilitate their journey toward the surface.

\subsection{The Flux Tube Paradigm}\label{sec:FluxTube}

Isolated magnetic flux tubes in the convection zone have a long history of study because they are convenient both analytically and computationally, and had until recently been able to sufficiently explain the observed properties of active regions. In most studies, they are given an `a priori' magnetic field strength and flux – it is taken for granted that the dynamo, via global or local processes, has somehow managed to create them -  and usually assume they have been formed at the bottom of the convection zone. There are two primary types of flux tube simulations – the thin flux tube approximation \cite[e.g.][]{Spruit1981,Caligarietal1995,Fanetal1993,Weber2011} and anelastic 2D/3D MHD simulations \cite[e.g.][]{EmonetMoreno1998,Fanetal2003,Fan2008,JouveBrun2009}. The thin flux tube approximation takes the flux tube as so thin that there is an instantaneous balance between the pressure outside the flux tube and the gas pressure plus magnetic pressure inside the flux tube. All physical quantities are taken as averages over the tube cross-section, and the tube is essentially a 1D string of mass elements, free to be accelerated in three dimensions by bulk forces in ideal MHD, including buoyancy, magnetic tension, the Coriolis force, and aerodynamic drag. In order to resolve the flux tube cross section, 2D or 3D MHD simulations are used. These simultaneously solve the full set of MHD equations, and in some cases, convection. But in order to meet the grid resolution typical of these models, they often have a flux too large for most active regions \cite[see][]{FanLRSP2021}.

Flux tube simulations have sought to explain the appearance of solar active regions, such as their latitude of emergence, tilting action in accordance with Joy's Law, and the general trend of a more coherent, less fragmented morphology for the leading polarity of an active region as depicted in Figure \ref{fig:ar11072}. For all of these examples, flux tube simulations have pointed toward the Coriolis force as the driver of the phenomenon. Consider three primary forces acting on a flux tube cross-section: a magnetic tension force directed toward the Sun's rotation axis, a buoyancy force directed radially outward, and the Coriolis force resulting from toroidal flow within the tube. As the tube traverses the convection zone, conservation of angular momentum drives a retrograde flow within the flux tube, resulting in a Coriolis force (as mentioned above) directed inward toward the rotation axis. When the magnetic field strength of the flux tube is strong (i.e. super-equipartition), the buoyancy force dominates and the flux tubes rise radially from their original latitude at the base of the convection zone. 
As the field strength of the flux tube decreases, the outward component of buoyancy diminishes compared to the inward component of the Coriolis force, and the resulting trajectory turns more poleward such that flux avoids emerging at lower latitudes  \cite[e.g.][]{ChoudhuriGilman1987,Caligarietal1995}. A fourth force acting on the flux tube, the drag force, is stronger for flux tubes of lower magnetic flux. As a result, flux tubes with lower initial values of magnetic flux around $10^{20}-10^{21}$ Mx are able to rise more radially than those of $10^{22}$ Mx \cite[e.g.][]{ChoudhuriGilman1987,DSilvaChoudhuri1993,Fanetal1993}.

If portions of the flux tube remain anchored deeper down in the convection zone, it is found within thin flux tube simulations that the material near the apex of a rising loop will both expand and diverge (although still with net retrograde motion), leading to a Coriolis force induced tilting of the loop toward the equator \cite[]{DSilvaChoudhuri1993}. Following the Joy's Law trend, these simulations also show an increasing tilt of the flux tube legs
with increasing latitude of emergence \cite[e.g.][]{DSilvaChoudhuri1993,Caligarietal1995}. This is expected if the Coriolis force is responsible for the tilting action, as the Coriolis force is proportional to sine(latitude). Additionally, the tilt angle is found to increase with increasing magnetic flux \cite[]{Fanetal1994}. Within thin flux tube simulations, the retrograde plasma motion near the flux tube apex contributes to a stronger magnetic field strength in the leading leg (in the direction of solar rotation) compared to the following leg \cite[e.g.][]{Fanetal1993,Fanetal1994}. It is noted that plasma is evacuated out of the leading flux tube leg into the following leg. Owing to the condition of pressure balance between the flux tube and its surroundings ($P_{i}+P_{b}=P_{e}$), this results in a stronger magnetic field strength for the leading side of the loop compared to the following \cite[]{Fanetal1993}. Here it is important to highlight that idealized flux tube simulations of all varieties are very efficient at conserving angular momentum \cite[e.g.][]{Fan2008,JouveBrun2009,Weber2011}, yet studies utilizing local helioseismology rules out the presence of retrograde flows on the order of $~100$ m/s in favor of flows not exceeding $\sim15$ m/s \cite{Birchetal2013}. In comparison, the buoyantly rising magnetic structures within the 3D convective dynamo simulations of \cite{Nelsonetal2014} are weakly retrograde and are actually prograde within the simulations of \cite{FanFang2014}. Within these 3D convective dynamo simulations, and perhaps within the Sun itself, flux emergence processes may deviate more from the `idealized' flux tube paradigm than originally thought.

Work has been done to study the twist of flux tube magnetic field lines in 2D and 3D MHD simulations. This body of work shows that if the magnetic field is not twisted enough along the flux tube axis, the flux tube tends to break apart and lose coherence as it rises \citep[see review by][]{FanLRSP2021}, although a curvature in the flux tube can partially mitigate this \citep[e.g.][]{Martinezetal2015}. Essentially, a minimum magnetic field twist rate (i.e. angular rotation of the magnetic field lines along the flux tube axis) is needed to counteract vorticity generation in the surrounding plasma caused by the buoyancy gradient across the flux tube's cross section. It is observed that active regions have a preferred helical twist 
of the magnetic field that is left-handed in the Northern hemisphere and right-handed in the Southern hemisphere \citep{Pevtsovetal2003}. But, \cite{Fan2008} finds the tilt of the rising flux tube ends up in the wrong direction if the twist is of the preferred hemispheric sign and strong enough to maintain coherence of the flux tube. If the twist of the field lines is reversed in handedness, the tilt angle is of the correct sign. Reducing the magnetic field twist per unit length also solves the hemisphere tilt problem, but then the tube becomes less coherent and looses more flux as it rises.

The flux tube simulations mentioned previously in this section (\ref{sec:FluxTube}) do not consider the impact of convection on the evolution of rising magnetism. However, it is absolutely clear that convection modulates flux emergence when it is included, provided that the magnetic field is not substantially super-equipartition \cite[e.g.][]{Fanetal2003,JouveBrun2009,Jouveetal2013,Weber2011,Weber2013}. Convective motions and magnetic buoyancy work in concert to promote flux emergence. Convection destabilizes the tube at the base of the convection zone, forcing parts to rise. As the tube bends, mass drains down the tube legs, making the apex less dense than portions deeper down, and that part of the tube also rises buoyantly. This, in combination with convective upflows, help the flux tube to rise toward the surface, while convective downflows can pin parts of the flux tube in deeper layers. By embedding thin flux tubes in a rotating spherical shell of solar-like convection, \cite{Weber2013} perform a statistical study to investigate how convection impacts flux tube properties that can be compared to solar active regions. Taking all their results into consideration, they attempt to constrain the as-of-yet unknown dynamo-generated magnetic field strength of active-region scale flux tubes. They find that tubes with initial field strength $\ge40$ kG are good candidates for the progenitors of large ($10^{21}-10^{22}$) Mx solar active regions.

In particular, \cite{Weber2013} find that convection tends to increase the Joy's Law trend, especially for mid-field-strength flux tubes of 40-50 kG. These flux tubes also take the longest time to rise due to the competing interplay of buoyancy and drag from surrounding turbulent flows. By `increasing the Joy's Law trend', the authors refer to a systematic effect that the addition of solar-like giant cell convection tends to boost the tilt angle at the same emergence latitude compared to simulations not subject to a convective velocity field. This is attributed, in part, to the associated kinetic helicity within the upflows. Taking all of the simulations together for tubes initiated $\pm40^{\circ}$ degrees around the equator with a magnetic flux of $10^{20}-10^{22}$ Mx and initial field strengths of 15-100 kG, the distribution of tilt angles peaks around $10^{\circ}$ degrees. This is in good agreement with the active region observations of \cite{Howard1996} and \cite{StenfloKosovichev2012}. Furthermore, similarly peaked tilt angle distributions are found for the buoyantly rising, dynamo-generated loops from the 3D convective dynamo simulations of \cite{Nelsonetal2014} and \cite{FanFang2014}. Perhaps this is indicative of similar processes at work in both these convective dynamo simulations and the thin flux tube simulations of \cite{Weber2013} - it is the turbulent, helical motion of convective upflows and the dynamics of the rising flux bundles themselves that contribute to the tilt angles extracted here.


\subsection{Beyond Idealized Flux Tubes} \label{sec:BeyondFluxTubes}
In section \ref{sec:FluxTube}, we introduced the idealized flux tube paradigm to describe the transport of magnetism from the deep interior toward the surface. In section \ref{sec:formB}, we noted that buoyantly rising magnetic structures have been found to arise from simulations of extend magnetic layers and form within wreaths of magnetism generated by dynamo action. In these latter two examples of MHD simulations, the buoyantly rising magnetism is \emph{not} in the form of idealized, magnetically isolated magnetic flux tubes. While simulations of idealized flux tubes are able to reproduce some properties of solar active region observables (see Section \ref{sec:FluxTube}), it may be unlikely that flux bundles rise within the convection zone entirely isolated from other nearby magnetic flux structures or a background field. Here we review studies of flux emergence that go beyond idealized flux tubes in an anelastic interior.

It is recognized that the presence of a background magnetic field and reconnection occurring between various magnetic flux structures have implications for the flux tube's evolution and the complexity of active regions. For example, \cite{PintoBrun2013} introduce a twisted flux tube in a 3D spherical convection zone with an evolving background dynamo. In comparison to the purely hydrodynamic case of \cite{JouveBrun2009}, the presence of the background magnetic field introduces a `drag' on the tube as it rises which is dependent on the orientation of the flux tube's magnetic field with respect to the background field. In particular, flux tubes with one sign of twist seem to rise faster than the ones possessing the opposite sign. The favored handedness then depends on the preferred magnetic helicity sign of the dynamo field.
By embedding a twisted toroidal flux tube in an effectively poloidal background magnetic field, \cite{Maneketal2018,Maneketal2021,Maneketal2022} show that a particular sign of twist increases the likelihood of a flux tube's rise and aligns with solar hemispheric helicity rules of active regions. Indeed, as mentioned in Section \ref{sec:obs}, observations show a tendency for active regions to possess a negative helicity in the Northern hemisphere and a positive one in the Southern hemisphere, although this is not a strict rule and only obeyed by only about $60\%$ of active regions \citep{Pevtsovetal2014}.

Beyond the interactions between buoyant concentrations of magnetic field and the dynamo-generated smaller-scale fields, it has also been argued that the reconnections between multiple buoyantly rising structures could have strong consequences on emerging regions. In particular, these reconnections can be at the origin of complex active regions, with strongly sheared polarity inversion lines and patches of positive and negative magnetic helicity, indicating a high potential for flaring activity. 
Simulations of such processes were conducted initially by \cite{Lintonetal2001} in a Cartesian geometry and then by \cite{Jouveetal2018} in a spherical shell including convection. In the latter, it was found that flux tubes with the same sign of axial field and same twist could merge to produce a single active region with a complicated structure and non-neutralized radial currents which could make these regions more likely to produce flares. Fully compressible calculations by \cite{Toriumietal2014} were also performed to explore the possibility that the intense flare-productive active region NOAA 11158 could be the product of interacting buoyant magnetic structures. Considering flux tubes isolated from the ambient dynamo field and independently rising to the solar photosphere is thus probably too simplistic.

The global models which simulate the interactions between convective motions, large-scale flows and more-or-less idealized, isolated magnetic flux tubes do not treat the upper-most layers of the convection zone and thus do not model the photospheric emergence. Firstly, the thin flux-tube approximation loses its validity above $\sim$$0.98R_\odot$, owing to the expansion of the tube apex to maintain pressure balance, to the extent that the tube radius becomes comparable with the local pressure scale height.  Secondly, the anelastic approximation also breaks down close to the photosphere where Mach number becomes order unity.
At this point, as a caution to the reader, we have to remember that the outcomes from these computational simulations serve only as touchstones for comparison to active regions. Direct comparisons between the properties of observed active regions and the results of thin flux tube simulations and the magnetic bipolar structures produced at the top of the computational domain of 3D anelastic simulations may be misleading.  

Compressible simulations including radiative transfer more closely approach the physics occurring at the top of the convection zone. These simulations aim at understanding how buoyant magnetic structures would make their way through the huge gradients of density and temperature  in this region. This work first started with \cite{Cheungetal2010} who used the MURaM code to simulate the photospheric emergence of a highly twisted torus placed at the base of the computational domain (around 7 Mm below), which was then gently brought towards the surface by an imposed radial velocity field of 1 km/s. This work was then extended to investigate the effects of less structured magnetic fields introduced at the bottom of the domain. In particular, \cite{Chenetal2017} used the flux concentrations produced by the convective dynamo simulations of \cite{FanFang2014} as an input, with a significant rescaling of the magnetic flux contained in these concentrations to have values at the photosphere compatible with typical active regions. Subsequently, an active-region-like structure was formed. 

Another example employing the MURaM code are the near-surface simulations of \cite{Birchetal2016}, where a torus of magnetic field, without twist, is introduced through the bottom boundary with varying speeds, up to the $\sim500$ m/s predicted rise speed of a thin flux tube \citep{Fan2008}. They found that  the strong diverging flows at the surface when the torus emerges are incompatible with observations, which do not show a significant diverging field.

Using the STAGGER code to model compressible, radiative MHD of the near-surface \cite{SteinNordlund2012} introduced an even less structured magnetic concentration by only imposing at the bottom boundary (at 20 Mm) a relatively weak untwisted uniform horizontal field of 1 kG. This field then rises towards the photosphere at the convective upflow speed and self-organizes into a bipolar active region with a coherent polarity and a more dispersed one. Several observational aspects like the rise speed, the absence of a strong retrograde flow, and the asymmetry between the polarities are interestingly reproduced in these simulations.
 
Other types of simulations of highly stratified turbulence also spontaneously produced magnetic flux concentrations resembling active regions, albeit not at the spatial or flux scales of real active regions, without the need to advect a well-defined magnetic structure at the bottom of the domain. It is the case for example of simulations by \cite{Brandenburgetal2013} and then \cite{Kapylaetal2016} where the Negative Magnetic Pressure Instability (NEMPI) mechanism is invoked to explain the spontaneous clumping of magnetic fields into a coherent structure. The most important ingredients in such simulations seem to be the strong density stratification and the large degree of turbulence. The formation of active regions following such a mechanism would then imply that they are produced in the subsurface layers of the Sun where both strong stratification and turbulence exist. If it turns out that such mechanisms are indeed at work in the Sun, this would completely revise our understanding of the flux emergence process and its origin. However, it is yet unclear how observed active region properties such as Joy's Law might be reproduced via the NEMPI mechanism.

The flux emergence process spans many orders of magnitude of density scale heights. Owing in part to this, it is difficult to get one singular simulation that tracks flux emergence from its generation by the dynamo to its interaction with the photosphere. As described above, some work has been done to `couple' flux emergence simulations of the deeper interior with those of a photosphere-like region. \cite{HottaIijima2020} performed the first radiative MHD simulation of a rising flux tube in a full convection zone, although without rotation, up to the photosphere. A 10 kG flux tube is introduced at 35 Mm below the top domain. Convection then modulates the flux tube, resulting in magnetic `roots' anchored in two downflows as deep as 80 Mm with a bipolar sunspot-like region forming at the apex of the now $\Omega$-shaped flux bundle. More realistic simulations like these, incorporating rotational effects, will make it increasingly straightforward to directly compare to, and interpret, solar observations.

\section{Statistical constraints supporting a passive active region emergence}
\label{sec:nearsurface}

The formation of each active region is unique. Simulations of active region emergence, especially those in 3D with appropriate active-region-scale magnetic flux, are currently too computationally expensive to build a statistically significant sample of flux emergence scenarios. \cite{Weber2011,Weber2013} have circumvented this somewhat by performing simulations of thin flux tubes embedded within a time-varying 3D convective velocity field (see Sec. \ref{sec:FluxTube}). This limiting factor makes it especially important to have a comprehensive sample of observed emerging active regions for comparison. 
Understanding the common properties of the emergence process is the only avenue to constrain the common physics behind flux emergence. There have been a number of statistical observational studies \cite[e.g.][]{Kommetal2008,KosovichevStenflo2008,Lekaetal2013,Birchetal2013,Barnesetal2014,McClintockNorton2016} on the formation of active regions, but in this paper we will focus on  the observed characteristics that can place direct constraints on the models. We refer to an `active' emergence as being guided by the magnetic field, and `passive' as being guided by the convection.

\subsection{Geometry of the flux tube}\label{sec:sepvel}
There is an apparent asymmetry in the east-west proper motions of the two main active region polarities as they emerge, with the leading polarity moving prograde faster than the following polarity moves retrograde \citep[e.g.][]{GilmanHoward1985,ChouWang1987}. Simulations have explained this as consistent with a geometrical asymmetry in the legs of an emerging flux tube, where the leg in the prograde direction is more tangentially oriented than the following leg which is more radial \citep[see for example Fig. 5 of][]{Jouveetal2013}. As this flux tube rises through the surface, the leading polarity moves more rapidly in the prograde direction than the following polarity in the retrograde direction. Modeled within the thin flux tube approach in particular, this asymmetry is due to the Coriolis force driving a counter-rotating motion of the tube plasma so that the summit of the loop moves retrograde relative to the legs \citep[e.g.][]{MorenoInsertisetal1994,Caligarietal1995}.

\begin{figure}[ht]%
\centering
\includegraphics[width=0.9\textwidth]{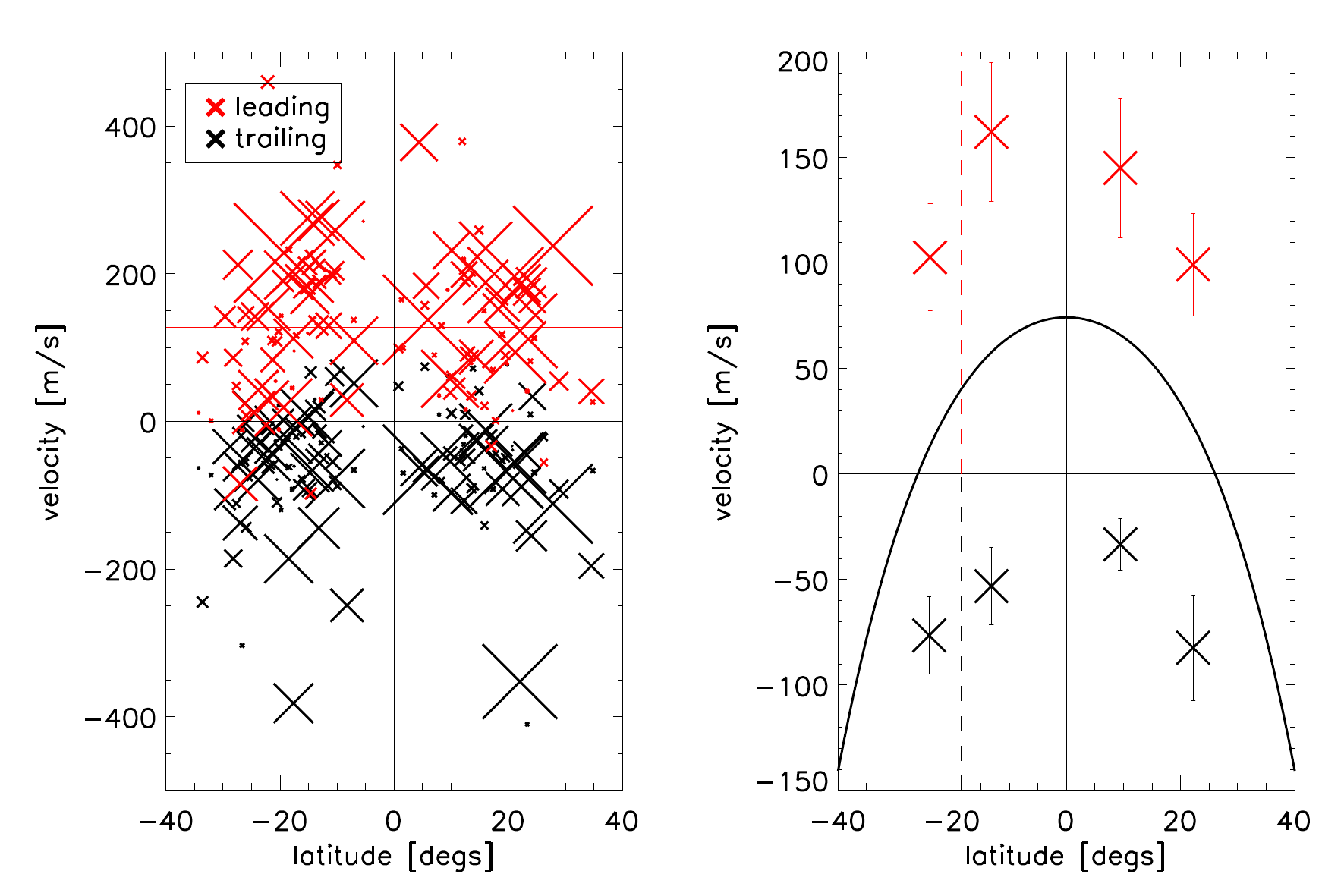}
\caption{Left: The mean east-west velocity relative to the Carrington rotation rate of the leading (red crosses) and trailing (black crosses) polarities over the first day after emergence for  117 active regions selected from the Solar Dynamics Observatory Helioseismic Emerging Active Regions Survey (SDO/HEARS; see Table A.1 in each of \cite{Schunkeretal2016,Schunkeretal2019} for a full list of active regions in SDO/HEARS and Appendix~\ref{app:earssepvel} for a list of active regions that were excluded). The size of the symbols represents the size of the active region (AR~11158 is the largest). The scatter is large; this emphasises the uniqueness of each active region emergence. The mean velocity of the leading polarity in the first day after emergence is $127 \pm 14$ms$^{-1}$ and the trailing polarity is $-61 \pm 10$ms$^{-1}$ . Right: The average velocities in bins of polewards and equatorwards latitudes divided by the median latitude (dashed vertical lines) of the EARs. The black curve shows the differential velocity of the surface plasma relative to the Carrington rotation rate. The uncertainties are given by the rms of the velocities in each bin, divided by the square root of the number of EARs in the bin. This figure is an updated version of Fig.~11 in \cite{Schunkeretal2016} where the average speed of the leading polarities was $121 \pm 22$ms$^{-1}$  and for the trailing polarities was $-70 \pm 13 $ms$^{-1}$. Full details of the method to measure the east-west polarity speeds are described in Section~7 of \cite{Schunkeretal2016}. }
\label{fig:sepvel}
\end{figure}

However, \cite{Schunkeretal2016} showed that while there is an apparent asymmetry of the leading and following polarity motion of the active region with respect to the Carrington rotation rate, this east-west motion is actually \emph{symmetric} with respect to the local plasma rotation speed. Here, in Fig.~\ref{fig:sepvel} we show the separation velocities for 117 active regions, an increased sample from what is shown in \cite{Schunkeretal2016}, further supporting the initial results. The average motion of the leading and following polarities in the first day after emergence is asymmetric with respect to the Carrington rotation rate (Fig.~\ref{fig:sepvel}, left), consistent with e.g. \cite{ChouWang1987}, where the mean east-west velocity of the leading polarity in the first day after emergence is $127 \pm 14$ms$^{-1}$ and the trailing polarity is $-61 \pm 10$ms$^{-1}$. However, we emphasise that \textit{the east-west motion of the polarities about the local plasma rotation speed is symmetric} (Fig.~\ref{fig:sepvel}, right). By embedding thin flux tubes within solar-like convection, \cite{Weber2013} show the average rotation speed of the center between the leading and following rising flux tube legs can approach the solar surface rotation speed, but only for strong flux tubes with initial field strengths of $\ge60$~kG. Although, due to strong conservation of angular momentum within the rising loop, the plasma flow at the apex of the loop is substantially retrograde (see also Sec. \ref{sec:FluxTube}) beyond what is detected by observations \citep[e.g.][]{Birchetal2013}.
  
Based upon these outcomes from observations and simulations, we suggest that any constraints placed on models of emerging flux tubes with geometrically asymmetric legs should be carefully reconsidered. Care should also be taken when choosing the particular reference frame to study their apparent motion asymmetries.

\subsection{Rise speed of the flux tube} \label{sec:risespeed}
In the absence of convection, idealized thin flux tube simulations show an upward rise speed of about 500~m/s at about 20-30~Mm below the surface \citep[e.g.][]{Caligarietal1995}. To understand how this would manifest at the surface of the Sun, \cite{Birchetal2016} inserted a torus of magnetic flux through the bottom boundary of a three-dimensional fully convective, near-surface simulation with a rise speed of 500~m/s. This simulation produced a strong outflow at the surface (about 400~m/s) as it emerged. Observations of the surface flows during an emergence on the Sun do not show such a strong outflow signature, but rather flow velocities that are consistent with a rise-speed less than $\approx100$~m/s, typical of convective upflows in the near-surface layers. In agreement with the observations, a flux tube that emerges naturally from a depth of 50~Mm  within the radiative MHD simulations of \citep{HottaIijima2020} does not produce any significant outflow at the surface. However, the rise speed of the flux tube is 250~m/s. This calls into question the traditional, idealized flux tube picture and suggests that the convection has an influence on the near-surface emergence process, but does not exclude thin flux tubes which may rise from the base of the convection zone with a slower speed.

\cite{HottaIijima2020} suggest that the reason their flux tube forms such a convincing active region structure is because it is initially placed across two coherent downflow regions. The downflows effectively pin the ends of the flux tube down, so that the centre emerges as a loop, implying that the influence of convection extends down to where flux tubes lie, and even instigate the emergence process, supporting some of the global models. Such correlations between rising (sinking) parts of flux tubes and upflows (downflows) were already observed in models of emergence in the bulk of the convection zone \citep[e.g.][]{Fanetal2003,Weber2011,Weber2013} and in rising flux bundles generated within 3D convective dynamo simulations \citep{Nelsonetal2011,Nelsonetal2013,FanFang2014}. But, the work of \cite{HottaIijima2020} shows that this interplay could also happen near the photosphere and highlights the potential importance of convective motions to bring the observed magnetic structures to the surface. 


\subsection{Onset of Joy's Law}\label{sec:joy}

Joy's Law is the observed tendency of the leading polarity in predominantly east-west aligned active regions to be slightly closer to the equator than the following polarity. The angle these polarities make relative to the east-west direction is called the tilt angle, and it increases with the latitude of the active region, strongly suggesting that Joy's Law has its origins in the Coriolis force. In some mean-field dynamo models, Joy's Law is an important characteristic where the tilt angle acts as a non-linear feedback mechanism  \citep[e.g.][]{Cameronetal2010}. 

Within the idealized flux tube paradigm, plasma near the rising flux tube apex will expand and diverge. This results in a Coriolis force-induced tilt of the tube axis in the sense of Joy's Law that increases with latitude (see Sec. \ref{sec:FluxTube}). In this picture, the tilt angle should be present at the time of emergence. The tilt angle also depends on the flux and field strength of the magnetism \citep[e.g.][]{Fanetal1994}. A larger magnetic flux $\Phi$ increases the buoyancy of the tube, and therefore the rise speed and effect from the Coriolis force, such that the tilt angle $\alpha$ increases ($\uparrow\Phi\Rightarrow\uparrow\alpha$). But larger magnetic field $B$ (for the same flux) increases the magnetic tension of the flux tube, which decreases the tilt angle due to the domination of tension over the Coriolis force \citep[$\uparrow B\Rightarrow\downarrow\alpha$, see also][]{Isik15}. \cite{Weber2013} show that incorporating the effects of time-varying giant cell convection systematically increases the tilt angles of rising flux tubes compared to the case without convection, but does not necessarily reproduce the tilt angle trends as found in \cite{Fanetal1994} (also see Sec. \ref{sec:FluxTube}). However, they do note that there is a larger spread in tilt angles at lower magnetic flux, as reported in some observations \citep[e.g.][]{WangSheeley1989,StenflowKosovichev2012}. Taken together, these simulation results show that the interplay of time-varying convection, the Coriolis force, magnetic tension, and buoyancy thus complicate trends in tilt angle.

\cite{Schunkeretal2020} measured the tilt angle of over 100 active region polarities throughout the emergence process, and found that on average, the polarities tend to emerge east-west aligned (i.e., with zero tilt), albeit with a large scatter, and the tilt angle develops during the emergence process. Moreover, \cite{Schunkeretal2020} found that the latitudinal dependence of the tilt angle arises only from the north-south motion of the polarities, and the east-west motion is only dependent on the amount of flux that has emerged.  They also found that there is no dependence of the tilt angle on the maximum magnetic field strength of the active region.
\cite{Schunkeretal2020} conclude that the observed Joy's Law trend is inconsistent with a rising flux tube that has an established, latitudinally dependent tilt angle as it rises to intersect with the photosphere.
We note that idealized thin flux tube models do not extend all the way to the surface (typically $\approx 0.98R_\odot$) where convection becomes important, however in simulations of coherent magnetic structures rising through the near-surface convection, the surface signature still reflects the orientation of the subsurface footpoints \citep[e.g.][]{Chenetal2017}.

Another possibility to explain Joy's Law is the conservation of magnetic helicity in a flux tube as it rises through the surface \citep[e.g.][]{BergerField1984,LongcopeKlapper1997}. The magnetic helicity is composed of the writhe, which measures the deformation of the flux tube axis, and the twist of the magnetic field lines about the axis. Ideally, the  magnetic helicity is a conserved quantity, and  changing one component necessarily requires a change in the other. In some simulations, the twist of the magnetic field about the axis of the flux tube is vital to it remaining coherent as it rises \citep[e.g.][]{Fanetal1998,Fan2008}. Within the thin flux tube context, it is shown that the writhe developed by the evolving flux tube can generate magnetic field twist \citep{LongcopeKlapper1997,FanGong2000}, but this alone is not enough to account for the observed twist of active regions \citep[see][]{FanLRSP2021}. Indeed, this relationship between the twist of the magnetic field and the writhe of the flux tube (related to the tilt of the active region) has been posited as a means to explore the link between `kink unstable' flux tubes and complex sunspot groups that have polarity orientations opposite to  Hale's Law  \citep[e.g.][]{Fuentesetal2003,FanLRSP2021}.
While there have been multiple studies of the helicity and twist of the surface magnetic field in active regions \citep[e.g.][and references therein]{Pevstovetal2014}, the relationship between the twist and writhe is still ambiguous. This is probably because observations do not have access to the full three-dimensional structure of the magnetic field above the surface, and only proxies for the twist and estimates of the helicity can be measured \citep[e.g.][]{Baumgartneretal2022}.

\begin{figure}[ht]%
\centering
\includegraphics[width=0.55\columnwidth]{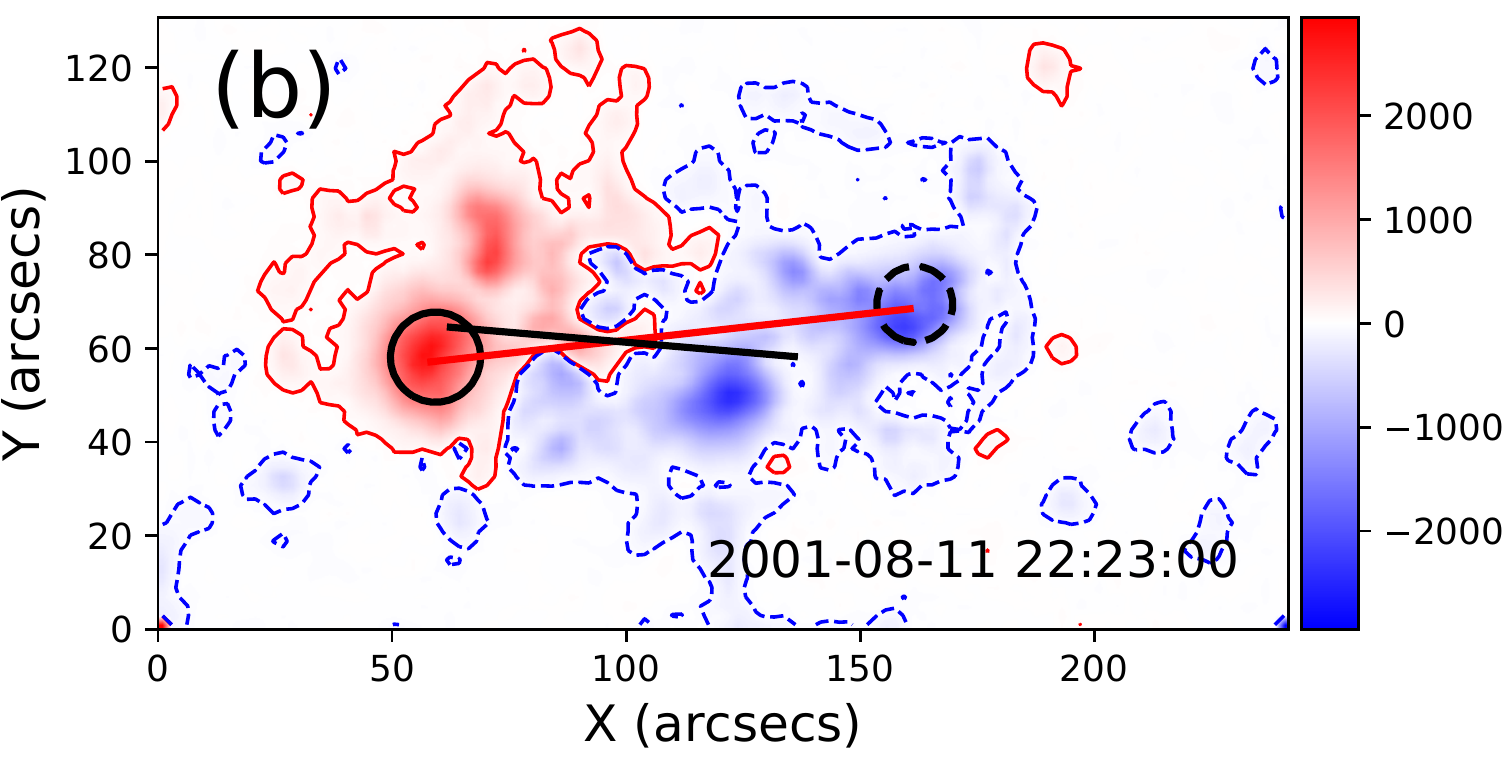}
\includegraphics[width=0.44\columnwidth]{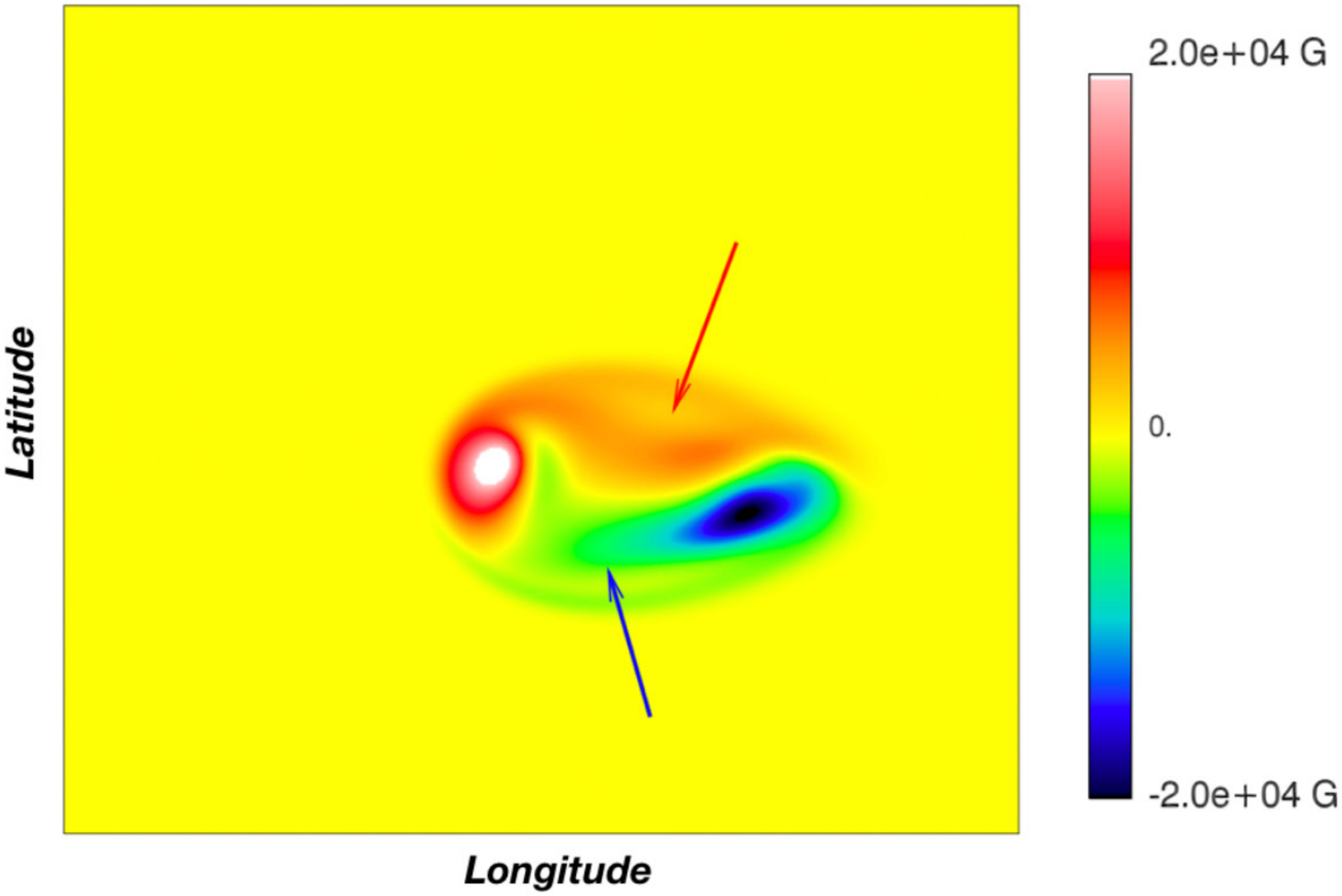}
 \caption{Left: Example of magnetic tongues observed by SOHO/MDI on the active region 9574. The blue and red shaded areas correspond to negative and positive values of the line-of-sight magnetic field (in units of gauss) and the black circles indicate the positions of the core flux of each polarity. Right: Example of magnetic tongues simulated in a global 3D MHD model of a twisted $\Omega$-shaped loop magnetic structure emerging close the top of the spherical shell (here $r=0.9 R_\odot$). Red and blue colours correspond to positive and negative radial magnetic fields and the arrows indicate the tongues of each polarity. \copyright AAS. Reproduced by permission from \cite{Poissonetal2020} (left panel) and \cite{Jouveetal2013} (right panel).} 
\label{fig:tongues}
\end{figure}

An interesting proxy for the global twist and writhe in active regions is the presence of so-called magnetic tongues (see Fig. \ref{fig:tongues}, left). These structures are due to the fact that the polarities of active regions appear elongated in line-of-sight magnetograms during their emergence \citep{LopezFuentesetal2000}. The elongation is thought to be produced by the line-of-sight projection of the azimuthal magnetic field at the peak of a twisted emerging flux-tube as it emerges through the surface. Thus, it is a proxy for the net twist of the active region flux tube, and coupled with the orientation  of the polarities (as a proxy for writhe), gives a constraint on the magnetic helicity brought to the photosphere by the emergence process \citep{Luonietal2011}. As the emergence proceeds, the tongues will vanish as the peak of the flux tube passes the surface and the legs of the flux tube remain. A less-biased measurement of the tilt angle will then be accessible. Magnetic tongues have also been reproduced in 3D MHD simulations where a twisted flux tube emerges through the deeper solar interior \citep[e.g.][and Fig. \ref{fig:tongues}, right]{Jouveetal2013} and closer to the photosphere \citep[e.g.][]{ArchontisHood2010,Cheungetal2010}. Again, a clear relationship can be established between the direction of elongation of tongues and the sign of the global active region twist, similarly to what is found in observations \citep{Poisson+22}.

The cycle-averaged tilt angles of sunspot groups show anti-correlation with the amplitude of the cycle \citep{Dasietal2010, Jiao+21}. A surface mechanism that explains this phenomenon is driven by inflows in the north-south direction towards active belts around the equator, effectively pushing the latitudinal separation of polarities of emerged active regions \citep{Jiang+10, Cameron12}. In the idealized thin flux tube picture, the same effect could be explained with enhanced cooling near the base of the convection zone where the strong toroidal flux is thought to be stored, which then pushes the onset of magnetic buoyancy to higher magnetic field strengths and thus magnetic tension, resulting in lower tilt angles at the surface \citep{Isik15}. The main assumption behind this mechanism is the reported systematic sound-speed reduction from solar minimum to maximum near the base of the convection zone \citep{BaldnerBasu08}. This implies stabilisation of flux tubes in the overshoot region, shifting the onset of magnetic buoyancy instability to higher field strengths to an extent that is consistent with the helioseismic inference \citep[see][for more details]{Isik15}.

An important task for future numerical simulations then is to decide to which extent Joy's law arises from (1) the latitude-dependent Coriolis force induced by diverging flows near the tube apex below the surface (ie, angular momentum conservation of horizontally diverging flows on a rotating fluid), (2) the interplay among twist, writhe and magnetic tension, and (3) the convective flows, which passively impose the tilt on flux bundles as they rise through the subsurface.

\section{Flux emergence and the solar dynamo}
\label{sec:dynamo} 

\subsection{Crucial role of active regions in global field reversals}

The magnetic flux that emerges through the photosphere in the form of bipolar magnetic regions is likely to play a key role in 
recycling the global magnetic field of the Sun. 
The idea that flux emergence as tilted bipolar active regions could play an active part in the dynamo process dates back to the 1960's with the seminal works of \cite{Babcock61} and \cite{Leighton69}. Indeed, in the so-called Babcock-Leighton (BL) model, the large-scale poloidal field owes its origin to the decay of sunspots at the photosphere. The leading polarity, closer to the equator, partially cancels with the opposite polarity in the other hemisphere, leaving a net flux to diffuse towards the pole to reverse the polar field of opposite sign. An important ingredient has then been added to this model - the large-scale meridional flow observed in the uppermost part of the convection zone \citep{Gizonetal2020}. These models including this flow are known as flux-transport dynamo models and are reviewed in another chapter of this book \citep{Hazraetal2023}. 

Recently, \cite{cameronmsch15} applied Stokes' theorem on the meridional plane of the Sun encompassing the convection zone to show that the net toroidal flux generated by differential rotation must come solely from the magnetic flux emerging at the surface. That surface flux mainly comes from the dipole moment contribution to the poloidal field of the Sun, which the tilted active regions eventually produce in the course of an activity cycle \citep{Cameronetal2018}. This theoretical finding highlighted the importance of flux emergence in solar and stellar dynamo processes. Indeed, similar analysis has been conducted by \cite{Jeffersetal2022} on two active K-dwarf stars followed by spectropolarimetry ($\epsilon$-Eridani and 51 Cygni A) where, similarly to the Sun, a balance is found between the generation of toroidal flux associated with the poloidal field threading through the stellar surfaces and the loss of magnetic flux associated with flux emergence. 

The latitudinal distribution and the tilt angle of emerging active regions thus seem of utmost importance in determining the global axial dipole of the Sun. As discussed in Sect.~\ref{sec:joy}, cycle-averaged tilt angle of sunspot groups are reported to show anti-correlation with the cycle strength \citep{Dasietal2010, Jiao+21}. This tendency has been interpreted as a manifestation of nonlinear saturation of the solar cycle. Accordingly, the effect works so as to limit further growth of the toroidal flux of the subsequent cycle. It does so by quenching the surface source for the global axial dipole moment through a lower average tilt angle of active regions. To account for the systematic effect, two physical mechanisms have been suggested: convergent flows towards emerged active regions, with the velocity depending on cycle strength \citep{Jiang+10}, or a deep-seated stabilisation of flux tubes by cooling, the extent of which depends on the toroidal magnetic flux \citep[][see Sec.~\ref{sec:joy}]{Isik15}.

It has to be noted here that despite observational evidence of the possible major role of surface fields and flux emergence in the dynamo process, all the global MHD dynamo models producing large-scale magnetic cycles today \citep[e.g.][]{Ghizaruetal2010,Nelsonetal2013,Kapylaetal2012,Augustsonetal2015,Hottaetal2016,Strugareketal2017} do not include the solar or stellar surface and do not produce starspots. This could indicate that the Sun is simply not operating in the same regime as 3D simulations. A more optimistic view would be to consider that a dynamo process relying on differential rotation and convection could still be active in the deep solar/stellar interior and that flux emergence would be an additional source of large-scale field not modelled yet in full 3D calculations.

\subsection{Incorporating flux emergence in Babcock-Leighton dynamo models}

Following the idea that flux emergence could play a key role in the dynamo process and that full 3D MHD global models do not yet capture all the characteristics of flux emergence, some works have been devoted to take prescriptions coming from 3D models of flux emergence and incorporate them into 2D mean-field Babcock-Leighton model. This was done, for example, in \cite{Jouveetal2010}. Here, the idea was to take into account the fact that flux tubes do not rise instantaneously to the surface (contrary to what is assumed in the standard BL model) and that the rise speed is a non-linear function of the magnetic field strength. They found that this small (but non-linear) delay in the rise time of flux tubes could produce long-term modulation of the cycle amplitude and phase.

Recently, the idea of combining the outcomes of 3D flux emergence simulations and 2D BL models has been used to produce new 3D flux-transport BL dynamo models, where active regions would be formed according to the toroidal field self-consistently created by the shearing of the poloidal field at the base of the convection zone \citep{YeatesMunoz2013, MieschDikpati2014, Kumaretal2019,Pipin2022, BekkiCameron2023}. These new models are particularly promising to study the role of active regions in the reversal of the polar magnetic field in the Sun and possibly other cool stars. Indeed, one of their advantages is that they are less prone to the caveat of 2D models of producing too much polar flux compared to observations. Moreover, in the last two references cited above, the non-linear feedback of the Lorentz force on the large-scale flows is taken into account and the impact of flux emergence on differential rotation and meridional flows can then be assessed. As further proof on the importance of active region tilt angles on the reversal of the Sun's poloidal field, \cite{KarakMiesch2017} find that introducing a tilt angle scatter around the Joy's Law trend in a 3D BL dynamo induces variability in the magnetic cycle, promoting grand maxima and minima. Many improvements still need to be implemented in these models, by incorporating statistics of flux emergence and characteristics of mean flows even closer to observations for example and possibly by implementing data assimilation techniques to construct predictive models for future solar activity \citep[see recent review by][on this subject]{Nandy2021}. Another improvement could also be to adapt these models to other stars with various emergence characteristics. Nonetheless simplified 3D BL models are already very valuable tools to be used before full 3D MHD models of spot-producing dynamos can be constructed.

\section{Flux emergence on other cool stars}
\label{sec:stars} 
\subsection{Some clues from observations}
\label{ssec:stars-obs}

Most stars with outer convection zones are capable of generating strong magnetism leading to starspots \citep[e.g.][]{Berdyugina2005,Strassmeier2009}.
The emergence of magnetic regions on other stars is not directly observable, however the strength and distribution of magnetic flux on the surface of stars can be inferred from  observations such as light-curve variability, (Zeeman-)Doppler imaging, and interferometry (see also van Saders et al. in this volume). This is only possible for stars significantly more active than the Sun, and it would not be possible to measure these properties treating the Sun as a star. 
The general trend for Sun-like stars is that for a given effective temperature, the unsigned surface magnetic flux 
increases with rotation rate until reaching a saturation point for faster rotators \citep[e.g.][]{Reinersetal2022}. There is also a preference for faster rotators to exhibit higher-latitude spots \citep[e.g.][]{Berdyugina2005}, but some rapidly rotating stars and fully convective M dwarfs can exhibit spots simultaneously at high and low latitudes \citep[e.g.][]{Barnesetal1998,Jeffersetal2002,Barnesetal2015,Davenportetal2015}.

It is then natural to wonder whether the observed trends of magnetic flux results from a link between the generation of the large-scale toroidal magnetic field and the bulk rotation rate.
Stellar rotation and effective temperature also affects the amplitude, vorticity, and turn-over time of the convection; in turn impacting the star's differential rotation (i.e. shear) profile \citep[see e.g.][and references therein]{BrunBrowning2017}. Some mean-field dynamos of the Sun incorporating a solar differential rotation profile find toroidal magnetic field generation with equatorward propagation near the tachocline \citep[e.g.][]{CharbonneauMacGregor1997,DikpatiCharbonneau1999,DikpatiGilman2001}. The tachocline is the name given to the region of radial and latitudinal shear at the interface between the solidly rotating radiative interior and the differentially rotating convection zone. These simulations emphasized that the tachocline is a key physical component in the solar dynamo mechanism. Yet, it is observed that even fully convective M dwarfs without tachoclines exhibit starspots and the so-called magnetic `activity-rotation correlation' \citep[e.g.][]{Reinersetal2014,WrightDrake2016,Reinersetal2022}. Further, some 3D convective dynamo models demonstrate that buoyantly rising magnetic flux structures can be generated within the bulk of the convection zone (see Sec. \ref{sec:formB}). With the recent emphasis on the role that convection plays (both local and mean flows) in active region emergence on the Sun (see also Sec. \ref{sec:nearsurface}), stars without tachoclines can offer some additional insights into how active-region-scale magnetism is manifested.

The variation in the 3D geometry of stellar photospheric magnetic fields poses another problem for numerical simulations of flux emergence. Zeeman-Doppler imaging of cool stars indicate that the magnetic energy in the toroidal component increases with the poloidal field for more active stars \citep{See+15}. Though with large scatter, the observational relation is steeper than one-to-one scaling for stars with masses above $0.5M_\odot$, with a power index of $1.25\pm 0.06$. 
The existence of a large amount of toroidal flux at the photosphere provides valuable constraints for the theory of magnetic flux emergence. Further analysis and interpretation by numerical simulations are needed to understand how such magnetic landscapes occur.

\subsection{Modelling the distribution of activity on stars: Hints from simulations}
\label{ssec:stars-mod}

\subsubsection{Active nests and longitudes}
\label{sssec:nests-longitudes}

We noted in Section \ref{ssec:stars-obs} that the unsigned surface magnetic flux increases with the rotation rate, for a given effective temperature. Whether this is due to increasing emergence frequency of active regions or larger sizes of individual active regions is unclear. These two scenarios do not exclude each other. An increased tendency for active regions to emerge near existing sites of emergence, known as active nests, is another possibility \citep[][see also van Saders et al. in this volume]{Isik+20}.

Observations indicate that the emergence of solar active features, including sunspots, coronal flares, and coronal streamers, are distributed inhomogeneously in longitude \citep[e.g.][]{Jetsuetal1997,BerdUso2003,Li2011}. Some other cool stars and young rapid rotators also exhibit these so-called `active longitudes' \citep[e.g.][]{Jarvinenetal2005,GarciaAlvarezetal2011,Luo2022}. The cause of active longitudes is still unknown, but a few theories have been put forward. One simple suggestion is that a long-lived localization of toroidal, amplified magnetic field at the base of the convection zone could spawn the onset of a magnetic buoyancy instability, promoting a series of rising flux loops \citep[e.g.][]{Ruzmaikin1998}. 
Similarly, the convective dynamo simulations of \cite{Nelsonetal2011,Nelsonetal2013} generate wreaths of magnetism within the convection zone that spawn buoyant bundles of flux when localized regions exceed a threshold field strength. Although, this effect is perhaps more closely related to the 'active nest' phenomenon described above.

Instead of relying on the localized enhancement of magnetic fields at particular longitudes, \cite{DikpatiGilman2005} show that MHD instabilities within a shallow water model of the tachocline can produce simultaneous variations in the tachocline thickness and tipping instabilities of the toroidal magnetic field there. A correlation between a `bulge' in the tachocline and a tipped toroidal band can force the magnetic field into the convection zone where it will rise buoyantly. \cite{WeberAL2013} present yet another alternate theory utilizing their thin flux tube simulations embedded in solar-like convection \citep{Weber2011,Weber2013}, which shows that active longitudes might also arise from the presence of rotationally aligned giant-cell convection. The simulations exhibit a pattern of flux emergence with longitudinal modes of low order and low-latitude alignment across the equator. Essentially, the extent of giant-cell upflows and the strong downflow boundaries form windows within which rising flux tubes can emerge. Although, \cite{WeberAL2013} use `active longitudes' to refer to a longitudinal alignment of flux emergence rather than repeated flux emergence at specific longitudes for multiple rotations. In reality, it is likely that both the amplification of localized magnetic fields and the effects of convective flows (which can also amplify localized fields) play a role in the active longitude and active nest phenomenon. 

Active longitudes have also been observed on stars in close binary systems \citep[e.g.][]{BerdyuginaTuominen1998,Berdyugina2005}. In this case tidal forcing was shown to affect the flux emergence patterns, leading to active longitudes on opposite sides of the star \citep{Holzwarth+msch03}. 
An exploration of the surface distribution of flux emergence for 
increasing stellar activity level has now become a necessity for physics-based 
numerical simulations, to better understand how stellar activity patterns 
scale with the activity level and the rotation rate. 

\subsubsection{Emergence latitudes and tilts}
\label{sssec:latitude-tilt}

Although highly simplified, simulations employing the thin flux tube approximation have been used as tools to explore the distribution of magnetic activity on stars with varying rotation rates \citep[][]{msch+96} and spectral types \citep[][]{Granzer+00}. These models once again point toward the importance of the rotationally-driven Coriolis force on flux tube dynamics (see also Sec. \ref{sec:FluxTube}). The existence of high-latitude and polar spots on stars with more rapid rotation and/or deeper convective envelopes can be explained by angular momentum conservation of a rising flux loop, leading to an internal retrograde flow. In the co-rotating frame, this effect would be experienced as an inward directed Coriolis force component towards the rotation axis, with a magnitude that increasingly dominates the radially outward buoyancy with more rapid rotation \citep[][]{msch+solanki92}.
The general trend is that beyond four times the solar rotation rate, a zone of avoidance forms around the equator, where no flux emergence occur \citep[][]{Isik+11,Isik+18}. 
In their simulations, the initial field strengths of toroidal flux tubes are assumed to be close to the analytical prediction of the onset of magnetic buoyancy instability. This limits the initial field strengths to the range 80-110 kG for the solar model with the initial tube location at the middle of the overshoot region below the convection zone. In a solar-type star rotating eight times faster, the range is in 150-350 kG, so that rotation stabilizes the tubes at a fixed field strength, owing mainly to angular momentum conservation \citep[][see Fig.~2]{Isik+18}. For rapidly rotating early K dwarfs and subgiants, the equatorial band of avoidance is somewhat widened in latitude, owing simply to the geometry of the convection zone boundaries: when the fractional depth of the convection zone increases (ie, towards cooler stars), the poleward-deflected tube apex can emerge at even higher latitudes \citep{Isik+11}.

These aforementioned simulations were based on the assumption that active-region producing flux tubes were formed near the base of the convection zone in the overshoot region, in the same way as for the idealised flux tubes in the Sun. It should be noted that in these studies \citep{Isik+11,Isik+18}, thin flux tubes rise in the presence of a differential rotation profile $\Delta\Omega$, which is kept constant with increasing stellar rotation rate. Taking notes from the simulations of \cite{Weber2011,Weber2013}, it is likely that incorporating turbulent, time-varying convective flows could modify these trends. 

\begin{figure}
\centering
\includegraphics[width=\columnwidth]{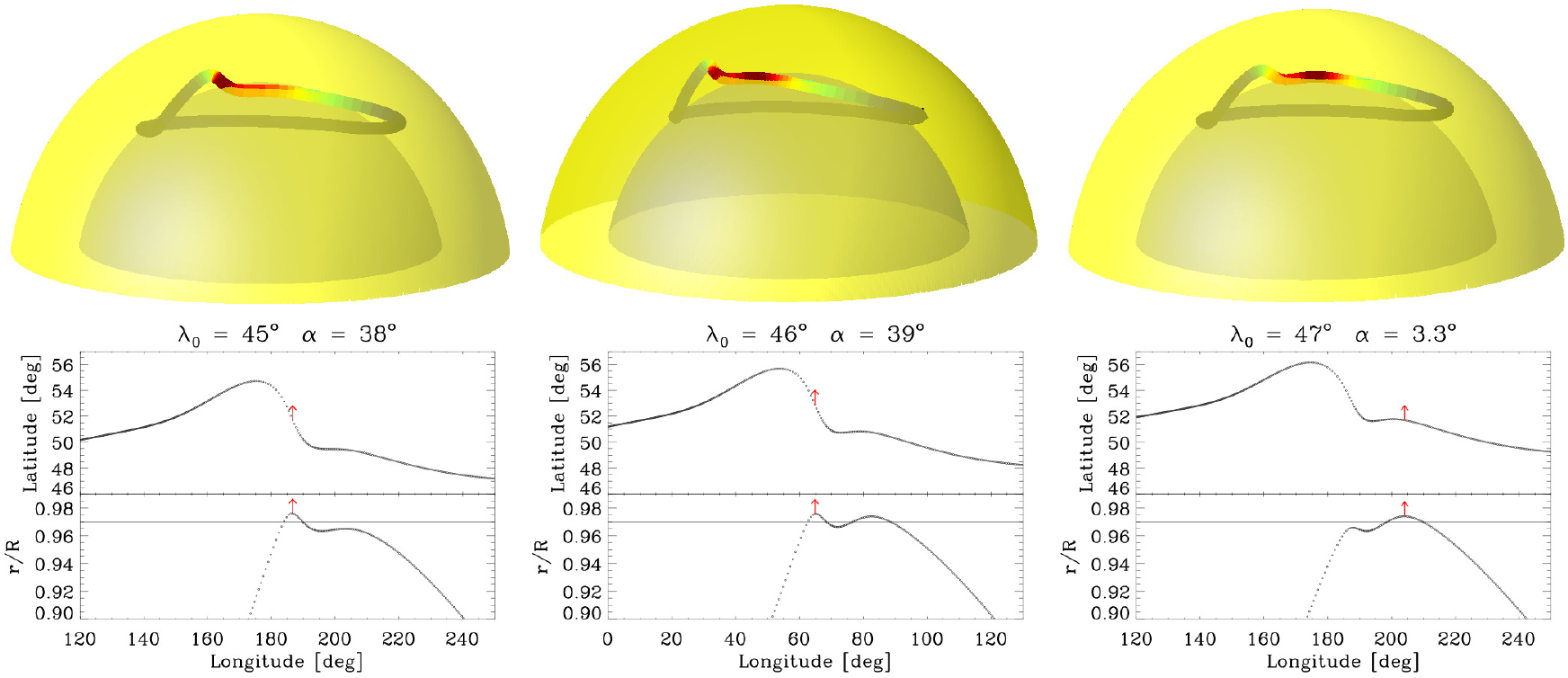}
\caption{Geometry of three emerging flux loops with initial latitudes $\lambda_0$ at the base of the convection zone and emergence tilt angles $\alpha$. Upper panels: The parts of the tube that are beneath the outer sphere ($0.97R_\odot$) are shaded in grey, whereas the emerged parts are brighter. The colours denote the cross-sectional tube radius (the redder the thicker). Lower panels: latitudinal and radial projections of the tubes. The horizontal line on the radial profile corresponds to the location of the outer sphere ($0.97R_\odot$), where $\alpha$ is measured from footpoint locations. The red arrows denote the apex of each tube.  I\c{s}{\i}k et al., A\&A, 620, A177 (2018), reproduced with permission © ESO.}
\label{fig:stellartilt}
\end{figure}

Thin flux tube simulations have also shown that the tilt angles near emergence generally increase with the rotation rate \citep{Isik+18}. This is consistent with the Coriolis acceleration along the tube apex being proportional to the local rotation rate. An increase of the tilt angle limits flux cancellation within the emerged bipolar regions and supports stronger fields to accumulate at the rotational poles \citep[see also][]{Isik+07}.\footnote{With the latitudinal distribution of emerging flux being confined to high latitudes, the stellar dynamo might not be dominated by the dipolar mode.} The tilt angles are not only larger in average than solar ones, but their variance is also larger, showing jumps at some emergence latitudes. Such a jump is demonstrated in Figure~\ref{fig:stellartilt}, which shows the detailed geometry of the flux tube apex starting close to $45^\circ$ latitude and emerging around $52^\circ$. When the initial latitude $\lambda_0$ is above $46^\circ$, the prograde part of the apex is intruded by a more east-west oriented and broader peak, leading to a tilt angle of about $3^\circ$. For $\lambda_0 < 46^\circ$, the large-tilt loop ($38^\circ$) emerges before the small-tilt loop. Possibly, such multiple-peaked adjacent loops emerge on active stars at certain latitudes, leading to complex active-region topologies with enhanced free energy deposits for the upper atmosphere.

M dwarfs with masses $\le0.35M_{\odot}$ are fully convective, and so lack a tachocline. Yet, in at least some ways, this magnetism is similar to that observed in Sun-like stars (see Sec. \ref{ssec:stars-obs}). \cite{Weber2016} embed the thin flux tube model within simulations of time-varying giant-cell convection to explore flux emergence trends in fully convective M dwarfs. Since there is no tachocline layer of shear, they introduce flux tubes at depths of $0.5R_\star$ and $0.75R_\star$ to sample the differing mean and local time-varying flows at each depth. A range of initial flux tube field strengths of 30-200 kG are chosen. On the lower end (30 kG), this encompasses magnetic fields that would not be too susceptible to suppression of their rise due to turbulent downflows. On the upper end (200 kG), this excludes field strengths above which the flux tubes would rise faster than they could plausibly be generated by large-scale convective eddies \citep{Browningetal2016}. Convection modulates the flux tubes as they rise, both promoting localized rising loops while suppressing the global rise of flux tubes (akin to magnetic pumping) for those initiated in the deeper interior at lower latitudes \citep[see also][]{Weberetal2017}. Within these simulations, a robust result is a tendency for flux tubes to rise parallel to the rotation axis (see Sec. \ref{sec:FluxTube} and the first paragraph in this section), leading to a preference of mid-to-high latitude flux emergence. However, low latitude flux emergence is found in special cases where the flux tubes are initiated closer to the surface and are of strong magnetic fields, or of weaker fields and rise through regions of prograde differential rotation near the equator.

\section{Moving forward}\label{sec:forward}

Active regions define the solar cycle, and in some models are an integral part of the transformation of the toroidal field to the poloidal field (Sec. \ref{sec:dynamo}). Understanding their deep-seated origins, formation and distribution will place tight constraints on their role in the solar dynamo and provide insights into these same processes in other cool stars. Typically, active-region-scale magnetism has been modeled as buoyantly rising, fibril tubes originating in the deep interior (Sec. \ref{sec:FluxTube}). New observations and simulations now suggest a shifting paradigm away from these idealized, isolated flux tubes toward a paradigm with a more complex, yet realistic, interplay between rising bundles of magnetism and their surroundings. 

Observations of surface magnetism demonstrate that active region flux emergence is a more `passive' process than was originally thought (Sec. \ref{sec:nearsurface}). The upward rise of the magnetism as detected near the surface is typical of convective upflows, placing much less of an emphasis on buoyancy in this region. However, it is not yet possible to say whether this influence of convection over buoyancy is confined only to the near-surface regions, or if it extends to the very beginning of the magnetism's rise. 

In idealized flux tube simulations, the Coriolis force leads to a geometrical asymmetry in the rising loop legs and a tilting action of these legs toward the equator. The former has been used as an explanation for why the leading active region polarity moves prograde faster than the following polarity moves retrograde. However, it is shown that the east-west motion of active regions is actually symmetric with respect to the local plasma rotation speed (Sec. \ref{sec:sepvel}). Further, the observed Joy's Law trend may not be consistent with the latitudinally-dependent tilt that the legs of a flux tube acquire as it rises through the convection zone (Sec. \ref{sec:joy}). The examples here and in the previous paragraph are observational evidence that flux emergence might be dominated less by buoyancy and the Coriolis force than was previously determined through flux tube models.

No global convective dynamo models have yet been able to produce starspots, partly because they do not include a realistic surface layer. Yet, we know that the surface distribution of emerging flux and the timing of their appearance is a key ingredient in Babcock-Leighton flux-transport dynamo models. Indeed, these incorporate ingredients self-consistently generated in global convective dynamo models such as differential rotation and meridional circulation. Also, they often assume that the primary region of magnetic field generation is at the tachocline. However, some convective dynamo simulations show that rising bundles of magnetism can be built within the bulk of the convection zone. At present, the exact generation region of active-region-scale magnetism and its strength is unknown. Learning more about the distribution of starspots across stellar photospheres for both Sun-like and fully convective stars may help to better constrain the interior source region of coherent magnetic structures. Knowing how the patterns of flux emergence vary as a function of stellar rotation and inferred surface differential rotation will also play a role in disentangling the imprints of rotation, mean flows, and shearing regions on the flux emergence pattern.

To fully understand the extent to which flux emergence is a passive process, more constraints from observations are still needed. But to understand what is happening below the surface, more simulations are critical. In particular, we suggest a strong emphasis to be placed on developing simulations that connect near-surface simulations with deeper flux emergence and dynamo models with fidelity. Further statistical analysis of solar active region emergence properties dependent on, for example, the extremes of magnetic flux and latitude, are also needed.

We conclude with some open questions regarding magnetic flux emergence in the Sun and other cool stars, raised by the observational and simulation landscape reviewed here, that we anticipate can be addressed in the next decade:

\begin{itemize}
    \item What properties of active region formation are driven primarily by the influence of convection?
    \item To what extent do the Coriolis force, convective flows, and tension, twist, and writhe of the magnetic field contribute to the observed Joy's law trend?
    \item What is the important physics that must be faithfully simulated to capture the observed statistical properties of emerging active regions?
    \item In our Sun, where is the primary region of generation for active-region-scale magnetism - the tachocline, near-surface, bulk of the convection zone, or some combination of these? 
    \item Can signatures of the underlying dynamo be found in patterns of magnetic activity (as reviewed here) on the photospheres of the Sun and other cool stars? 
\end{itemize}

\backmatter

\bmhead{Acknowledgements}
HS is the recipient of an Australian Research Council Future Fellowship Award (project number FT220100330) funded by the Australian Government and her contribution is partially funded by this grant. LJ acknowledges funding by the Institut Universitaire de France.

\section*{Declarations}


\begin{itemize}
\item \textbf{Conflict of interest:} The authors have no conflicts of interest to declare that are relevant to the content of this article.
\end{itemize}

\begin{appendices}
\section{Defining the duration of active region emergence}\label{app:earssepvel}

Here we describe how we computed the duration of the active region emergence process as shown in Fig.~\ref{fig:emergtime}.

\begin{figure}[ht]%
\centering
\includegraphics[width=0.99\columnwidth]{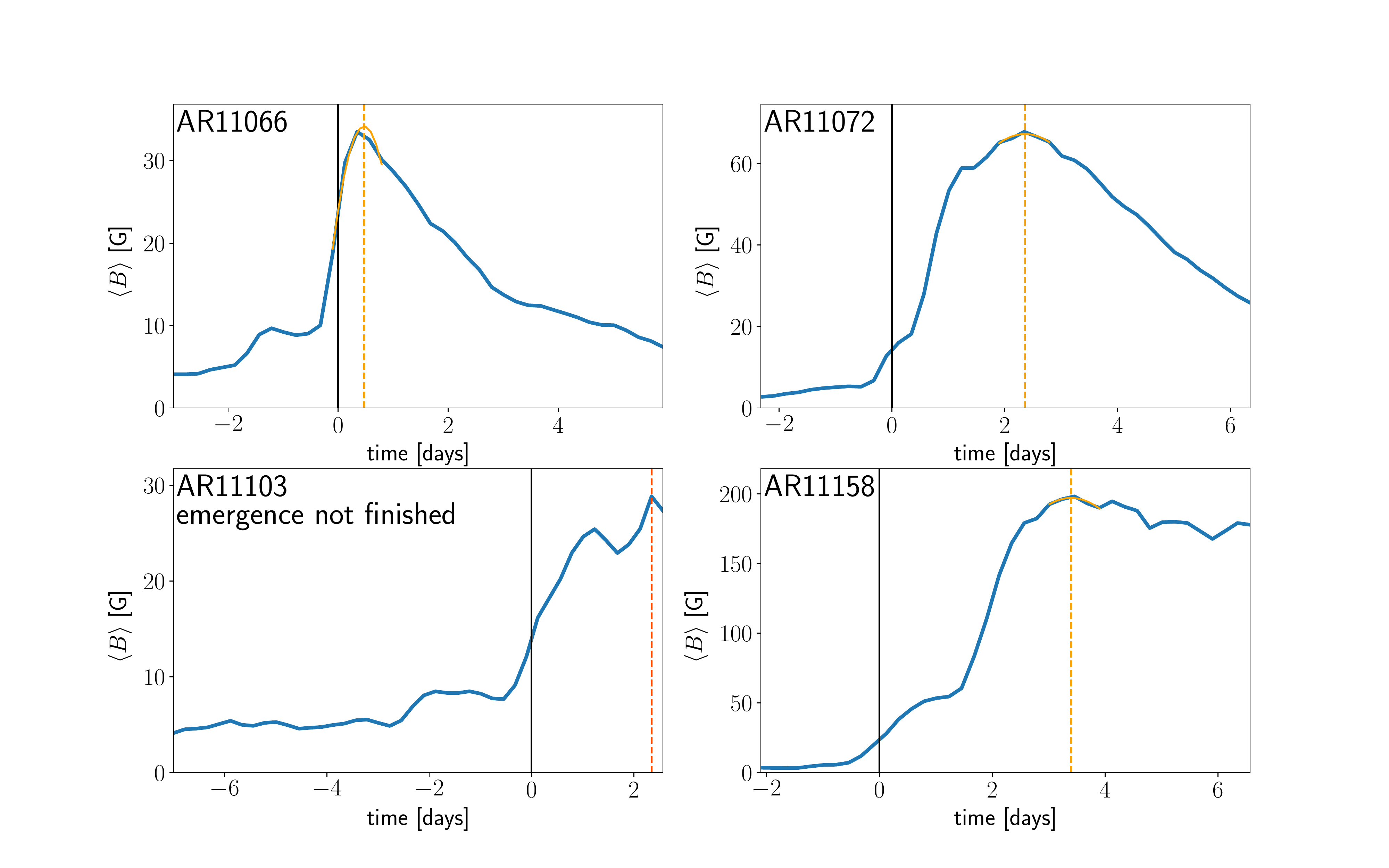}
\caption{Examples of the evolution of the averaged unsigned magnetic field within the central $49$~Mm radius, $\langle |B| \rangle$, for four emerging active regions relative to the emergence time  (blue). For AR11066, AR11072 and AR11158 the orange curve shows the least-squares fit of a quadratic to the peak of the curve and the dashed vertical line shows the time of the maximum value of the quadratic. Since the maximum value of $\langle |B| \rangle$ for AR11103 occurs within the last 16~hours of the time series, we exclude it from our sample.}
\label{fig:bev}
\end{figure}

Figure.~\ref{fig:bev} shows the evolution of the mean unsigned magnetic field, $\langle B \rangle$ within the central region of $49$~Mm radius the map. The emergence time is defined as being 10\% of the total magnetic flux 36~hours after the active region was officially named \citep[see ][ for more details]{Schunkeretal2016}. The maps are centred on the flux-weighted centre of the active region about the time of emergence \citep[for a detailed definition see ][]{Schunkeretal2016,Birchetal2016} for four example active regions from the Solar Dynamics Observatory Helioseismic Emerging Active Regions survey (SDO/HEARS) \citep{Schunkeretal2016}   which contains a total of 180 active regions.

The emergence of an active region ends when all of the magnetic field has appeared at the surface. We identified the time when the mean line-of-sight magnetic field $\langle B \rangle$ with a $5.3$~hour cadence was a maximum.
If the maximum occurred within three time intervals ($\approx 16$ hours) of the end of the time series it is difficult to assess whether it is still emerging (e.g. AR11103 in Fig.~\ref{fig:bev}), and so we exclude these regions (35 in total).
Otherwise, we fit a quadratic to the $\langle B \rangle$ values between 8 hours (two time intervals) before and 13 hours (three time intervals) after the time when $\langle B \rangle_\mathrm{max}$ occurred, and we defined the time of the maximum of the quadratic function as the end time of the emergence process. There is no physical basis for fitting a quadratic, only that we found it fit the peak reasonably well (see Fig.~\ref{fig:bev}). The duration of the emergence process is the difference between end of the emergence process and the emergence time.

List of 120 NOAA active region numbers included in Fig.~\ref{fig:sepvel}:
 11066, 11070, 11072, 11075, 11076, 11079, 11081, 11086, 11088, 11103, 11105, 11114, 11122, 11132, 11136, 11137, 11138, 11141, 11142, 11145, 11148, 11154, 11158, 11159, 11167, 11198, 11199, 11200, 11206, 11209, 11211, 11214, 11223, 11239, 11273, 11288, 11290, 11297, 11300, 11304, 11310, 11322, 11327, 11331, 11381, 11397, 11400, 11404, 11406, 11414, 11416, 11431, 11437, 11446, 11450, 11456, 11472, 11497, 11500, 11510, 11511, 11523, 11531, 11547, 11549, 11551, 11554, 11565, 11570, 11574, 11597, 11603, 11607, 11624, 11626, 11627, 11631, 11640, 11645, 11686, 11696, 11699, 11703, 11707, 11712, 11718, 11736, 11750, 11780, 11781, 11784, 11786, 11789, 11807, 11813, 11821, 11824, 11833, 11843, 11855, 11867, 11874, 11878, 11886, 11894, 11915, 11924, 11946, 11962, 11969, 11978, 11992, 12003, 12011, 12039, 12064, 12078, 12099, 12118, 12119.

\end{appendices}





\newcommand{\nat}{Nature} 
\newcommand{\apjl}{ApJL} 
\newcommand{\apj}{ApJ}
\newcommand{\aap}{A\&A} 
\newcommand{\aapr}{A\&A Rev.} 
\newcommand{\mnras}{MNRAS} 
\newcommand{\solphys}{Sol.~Phys.} 
\newcommand{\ssr}{Space.~Sci.~Rev.} 
\bibliography{chapter6}


\end{document}